\documentclass[12pt]{iopart}
\usepackage{epsfig}
\usepackage[footnotesize]{caption2}
\setcaptionmargin{5pt}

\begin{document}

\title[Event-by-Event Fluctuations at 40, 80, and 158 AGeV/c in Pb+Au Collisions]{Event-by-Event Fluctuations
at 40, 80, and 158 AGeV/c in Pb+Au Collisions}

\author{Hiroyuki Sako{\dag}\ and Harald Appelsh\"{a}user\
for the CERES/NA45 collaboration
}

\address{Gesellschaft f\"{u}r Schwerionenforschung (GSI),
64291 Darmstadt, Germany}

\ead{{\dag}\ h.sako@gsi.de}

\begin{abstract}
Event-by-event fluctuations of mean $p_{T}$ and net charge
in 40, 80, and 158~AGeV/c Pb+Au collisions are presented. The observed
dynamical mean $p_{T}$ fluctuations in central events of about 1~\% are
very similar to results from RHIC. The centrality dependence of mean
$p_{T}$ fluctuations at 158 AGeV/c shows a peculiar non-monotonic pattern
with respect to the extrapolation from the fluctuations measured in
p+p collisions. Dynamical net charge fluctuations smaller than the charge
conservation limit are observed. After correction for charge
conservation, the measured fluctuation signal in central events
is comparable to RHIC results.

\end{abstract}




\section{Introduction}
Event-by-event fluctuations of mean $p_{T}$ have been proposed as a
possible probe to search for the phase transition, especially the QCD
critical point via their non-monotonic variation
with control parameters such as beam energy and centrality~\cite{Stephanov:1999}.
On the other hand, the fluctuations can be
used to evaluate the extent of thermal equilibrium by comparison with the
extrapolated fluctuations from nucleon-nucleon collisions.
Suppressed net charge fluctuations have been proposed as a possible 
signal for the formation of a QGP, due to the smaller charge units of
deconfined (anti-) quarks compared to those of confined hadrons~\cite{Asakawa:2000,Jeon-Koch:2000}.
Comparisons with cascade models may clarify whether the observed
fluctuations are described by resonance gas models.

A detailed description of CERES experiment at the CERN-SPS can be found
elsewhere~\cite{Adamova:2002,ceres-ptfluc-2003}.
For the present analysis, momenta of charged
particle tracks produced in Pb+Au collisions at 40, 80, and 158 AGeV/c
have been measured with a cylindrical Time Projection Chamber. 
The data set comprises a wide range of collision centrality. 
The acceptance of the TPC is around mid-rapidity ($2.0<\eta<2.9$)
and covers $2\pi$ in azimuth (at 40~AGeV/c only $1.3\pi$, see also~\cite{ceres-ptfluc-2003}).
\section{Mean $p_{T}$ Fluctuations}
\begin{figure}
\begin{minipage}[t]{0.55\textwidth}
\begin{center}
\mbox{\epsfig{file=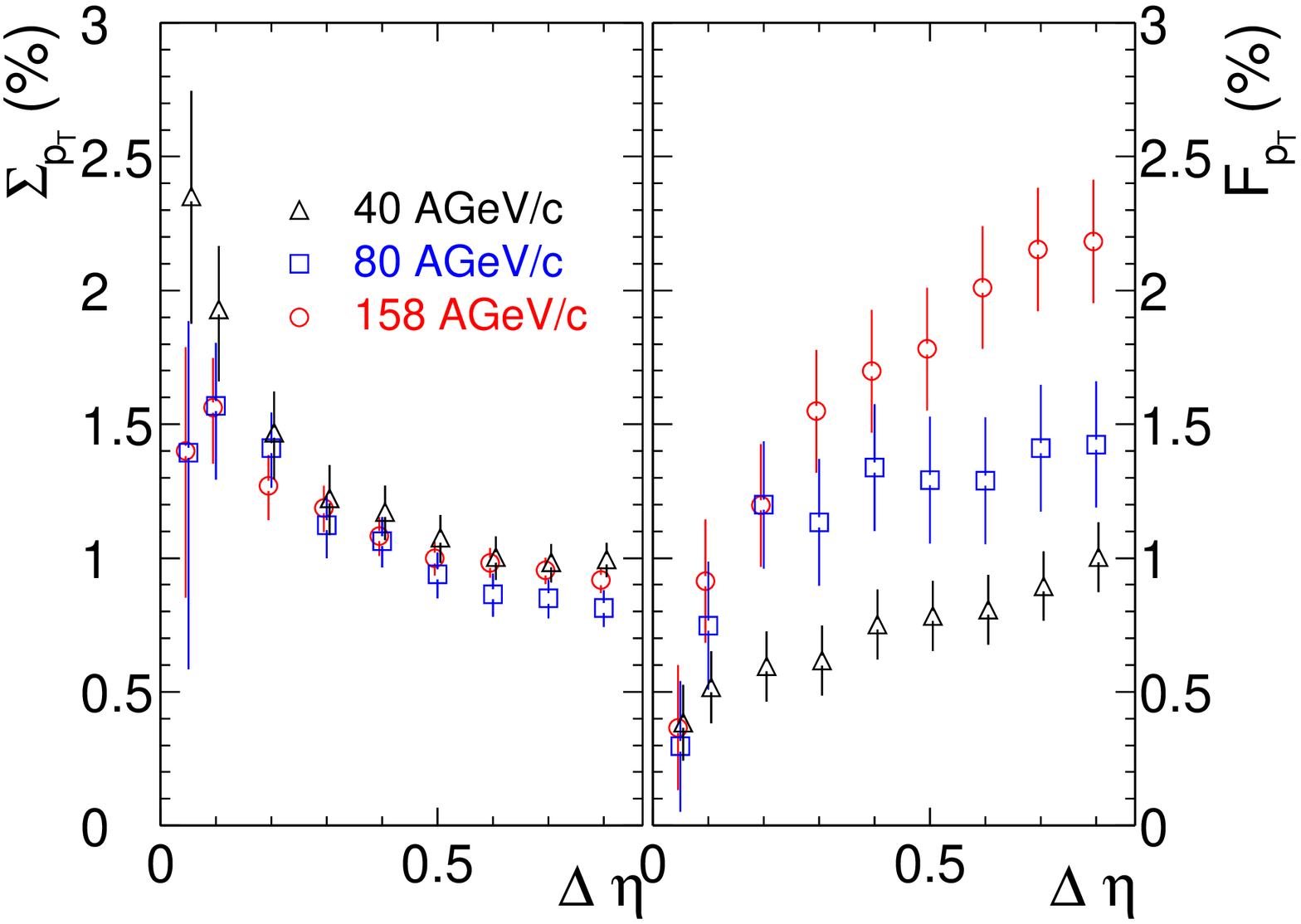,width=0.85\textwidth}}
\caption{$\Sigma_{p_{T}}$ (left) and $F_{p_{T}}$ (right) as a function
	of {$\Delta\eta$} at $0.1<p_{T}<1.5$ GeV/c in 6.5~\% central events.}
\label{fig:fluc-deta}
\end{center}
\end{minipage}
\begin{minipage}[t]{0.45\textwidth}
\begin{center}
\mbox{\epsfig{file=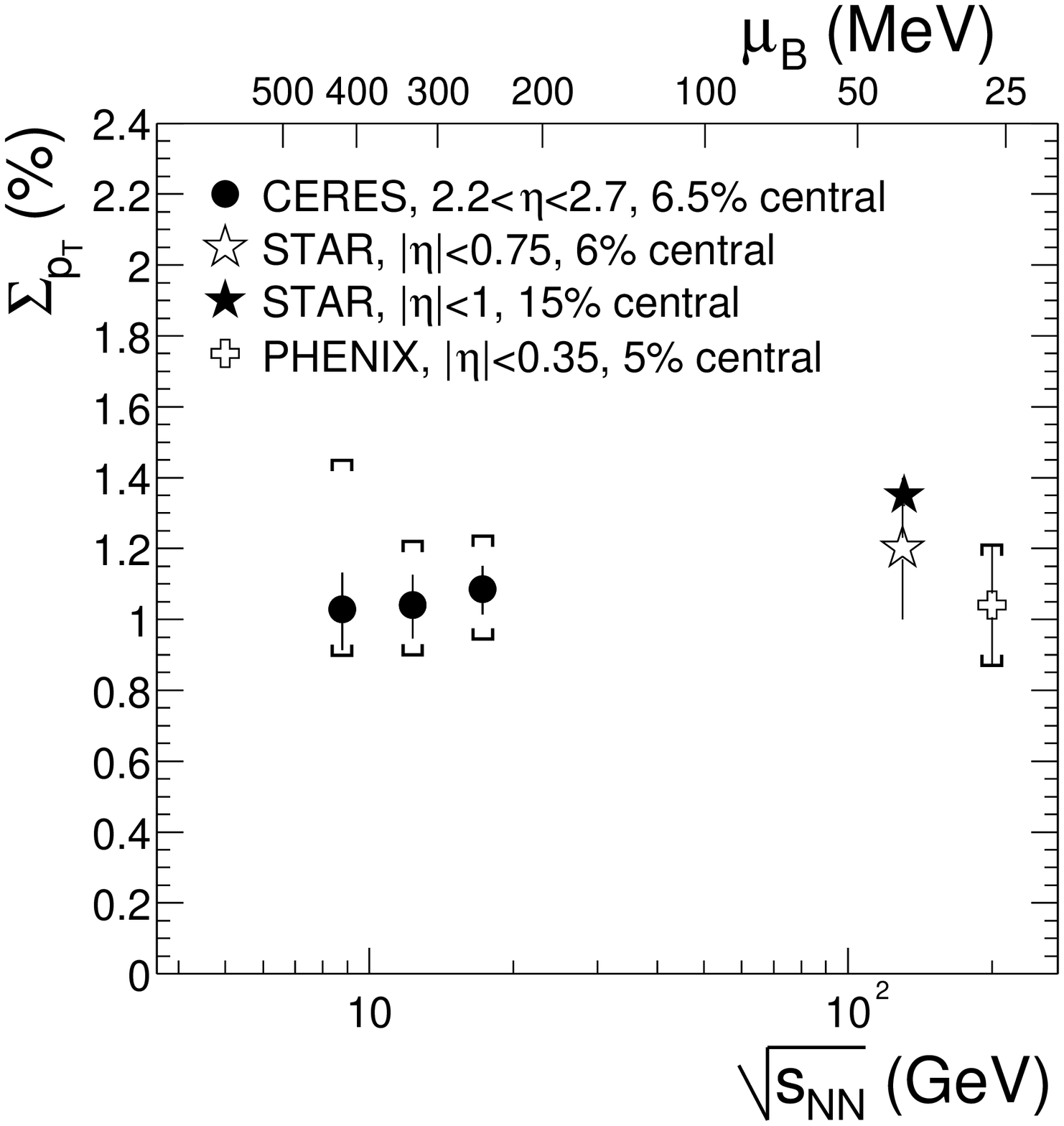,width=0.85\textwidth}}
\caption{$\Sigma_{p_{T}}$ as a function of {$\sqrt{s_{NN}}$}
      in central events.}
\label{fig:sdyns}
\end{center}
\end{minipage}
\end{figure}

Two measures are used to evaluate mean $p_{T}$
fluctuations. The $\Sigma_{p_{T}}$ measures
the dynamical fluctuation normalized with mean
$p_{T}$~\cite{ceres-ptfluc-2003}.
This measure is proportional to mean
covariance of all charged particle pairs per
event~\cite{Voloshin:1999}.
Another measure $F_{p_{T}}$~\cite{PHENIX-pt:2003} is defined as a
deviation from 1 of the ratio of the r.m.s.~of the event-by-event mean
$p_{T}$ distribution in real events to that in mixed events.
The $F_{p_{T}}$ is approximately proportional to
$\langle N\rangle \Sigma_{p_{T}}^{2}$, where $\langle N\rangle$ is the mean charged particle multiplicity
in the acceptance.
In Fig.~\ref{fig:fluc-deta}, $\Sigma_{p_{T}}$ has a saturation tendency at
the pseudo-rapidity interval ($\Delta \eta$) more than 0.4, whereas
$F_{p_{T}}$ steadily increases.
We therefore use $\Sigma_{p_{T}}$ to compare CERES results ($\Delta \eta =0.5$) with RHIC results ($\Delta \eta = 0.7 \sim 2$).

Fig.~\ref{fig:sdyns} shows $\Sigma_{p_{T}}$ as a function of
nucleon-nucleon center of mass energy ($\sqrt{s_{NN}}$) at
$2.2<\eta<2.7$ and $0.1<p_{T}<2$~GeV/c in 6.5~\% central events,
compared to RHIC data at $\sqrt{s_{NN}} = 130$~GeV~\cite{Voloshin:2001,STAR-pt:2003},
and $200$ GeV~\cite{PHENIX-pt:2003}. For comparison to RHIC data,
the CERES data are not corrected for short range momentum
correlations. The observed fluctuations at
SPS and at RHIC are similarly about 1~\%. No indication for
non-monotonic variation or enhanced fluctuations is observed. Note
that a predicted fluctuation at the critical point
is $\sim 2$\%~\cite{Stephanov:1999}, far above the measured values.




\begin{figure}
\begin{minipage}[t]{0.5\textwidth}
\begin{center}
\mbox{\epsfig{file=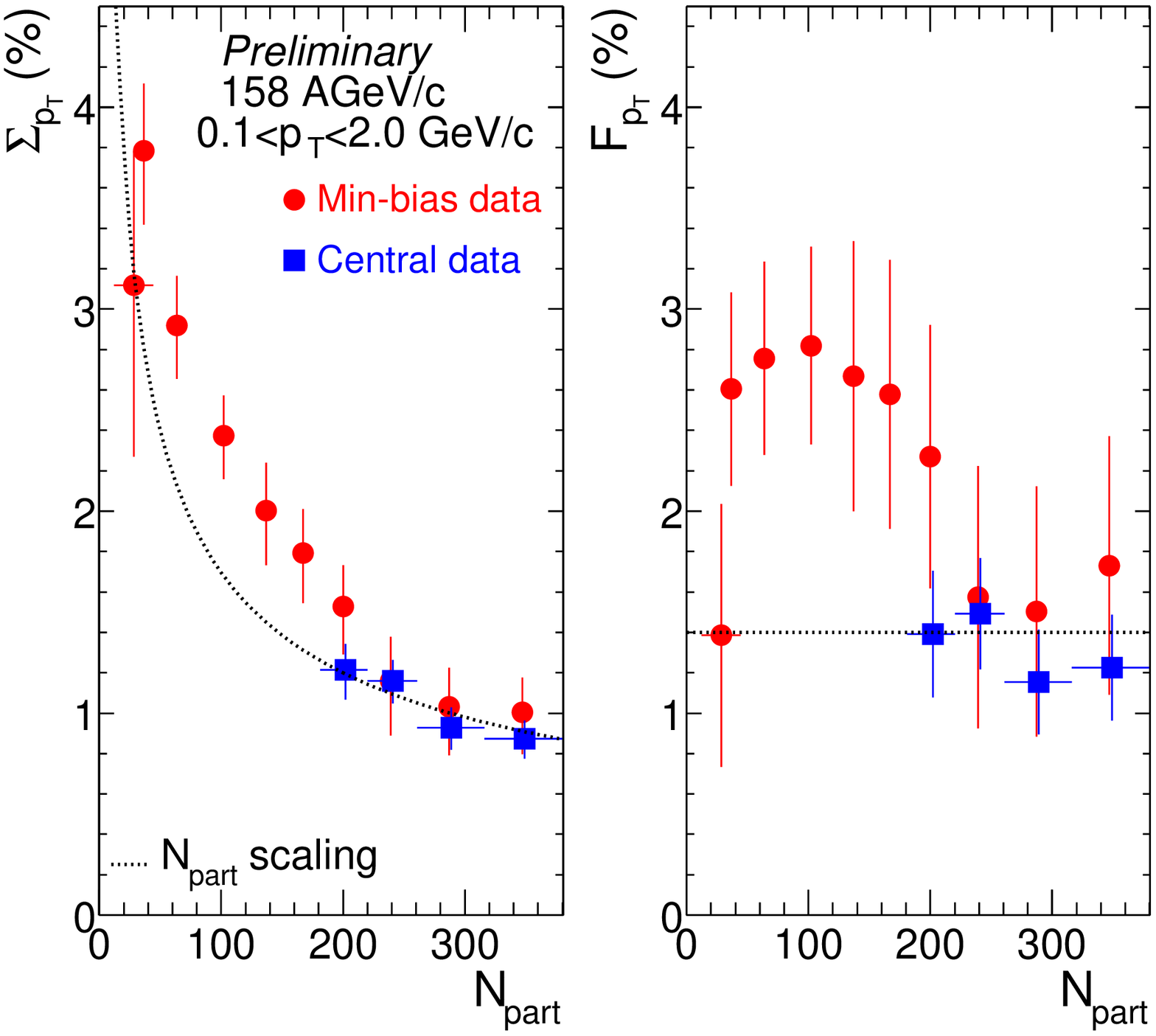,width=0.8\textwidth}}
\caption{$\Sigma_{p_{T}}$ (left) and $F_{p_{T}}$ (right) as a function of $N_{part}$
 at 158 AGeV/c.}
\label{fig:fptcent}
\end{center}
\end{minipage}
\begin{minipage}[t]{0.5\textwidth}
\begin{center}
\mbox{\epsfig{file=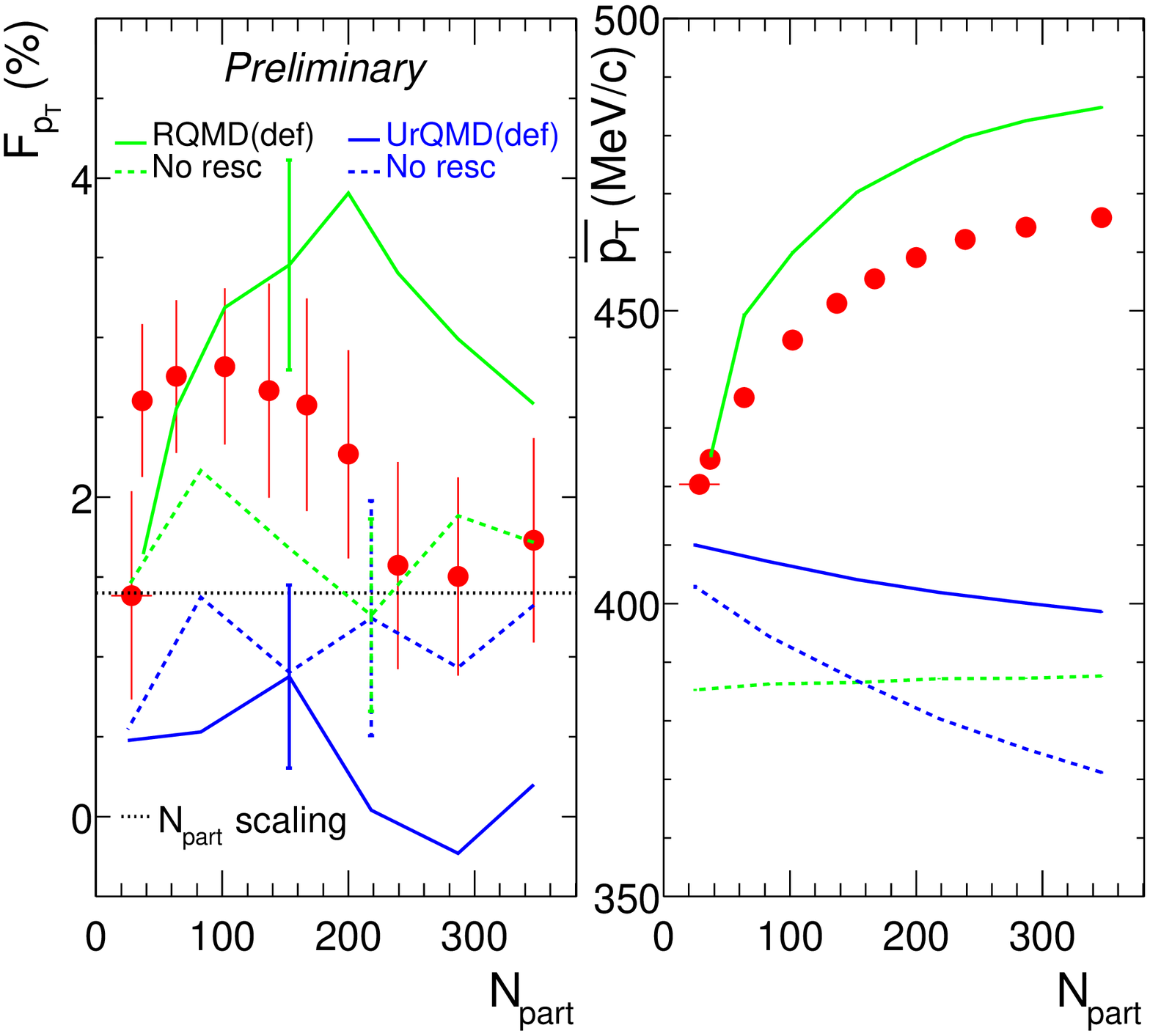,width=0.8\textwidth}}
\caption{$F_{p_{T}}$ (left) and $\overline{p_{T}}$
	(right) as a function of
	$N_{part}$ compared to RQMD and UrQMD at 158 AGeV/c.}
\label{fig:fptmpt}
\end{center}
\end{minipage}
\end{figure}

Fig.~\ref{fig:fptcent}
shows fluctuations corrected for short range
momentum correlations at 158 AGeV/c as a function of the number of
participants ($N_{part}$).
The fluctuations between the minimum-bias data set and
the central data set are consistent. As a baseline (dashed
lines) reflecting
the superposition of individual nucleon-nucleon collisions, we employ the
formula
$\Sigma_{p_{T}}^{AA}  = 12\% \cdot (\langle N_{part}\rangle/2)^{-1/2}$,
where the 12~\% is a measured fluctuation in p+p collisions at
ISR~\cite{braune_pp}, which corresponds to $F_{p_{T}} \simeq 1.4~\%$ in the
CERES acceptance.
Significant deviations from the baseline are
observed in semi-central events, whereas in peripheral events
(centrality $\geq  66\%$) and in central events (centrality $\leq 20\%$), the
fluctuations are consistent with the baseline. The maximum of
$F_{p_{T}}$ is about 2.8~\% at $N_{part}$ of $50 \sim 100$, corresponding to
$25\sim50~\%$ central events. The observed centrality dependence
as well as the increase of $F_{p_{T}}$ with the upper $p_{T}$ cut is
similar to the observations by NA49~\cite{NA49-pt:2003},
PHENIX~\cite{PHENIX-pt:2003}, and STAR~\cite{STAR-pt:2003}.

In Fig.~\ref{fig:fptmpt}, $F_{p_{T}}$ of RQMD and UrQMD without
rescattering is consistent with the baseline as expected.
With rescattering, RQMD reproduces qualitatively
both the non-monotonic dependence of fluctuations and the increasing
tendency of mean $p_{T}$, while UrQMD fails to describe both observables.


\section{Net charge fluctuations}
As a measure of net charge fluctuations, we use
$\nu_{dyn}$, representing dynamical fluctuations of the difference between
normalized multiplicity of positive particles and that of negative
particles~\cite{Pruneau:2002}. 
The measure is advantageous since
the correction for the global charge conservation is
additive, and no efficiency correction is necessary.

\begin{figure}
\begin{minipage}[t]{0.54\textwidth}
\begin{center}
\mbox{\epsfig{file=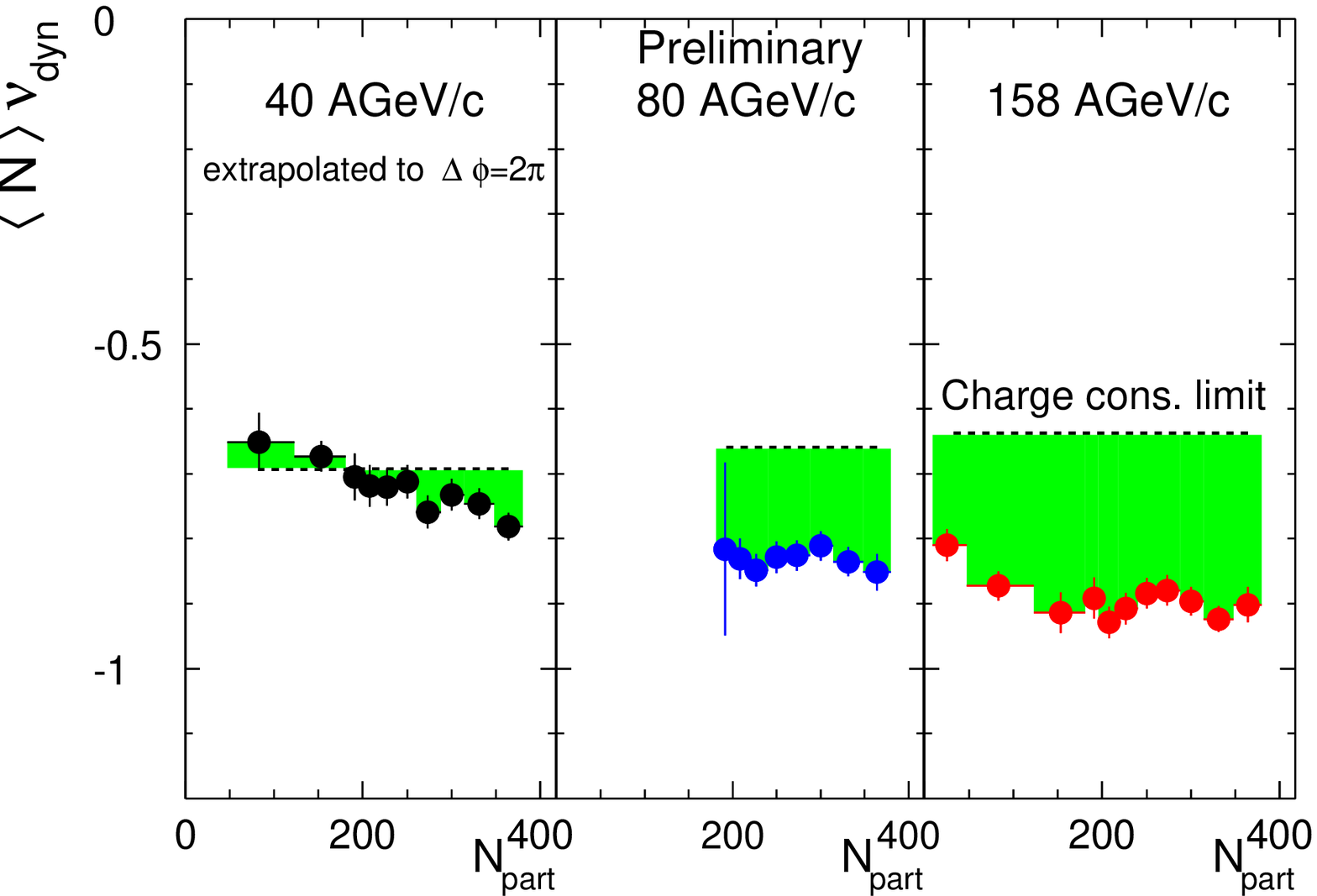,width=0.85\textwidth}}
\caption{$\langle N\rangle \nu_{dyn}$ as a function of $N_{part}$ at
      $2.05<\eta<2.85$ and $0.1<p_{T}<2.5$ GeV/c.}
\label{Fig:nvdyn}
\end{center}
\end{minipage}
\begin{minipage}[t]{0.46\textwidth}
\begin{center}
\mbox{\epsfig{file=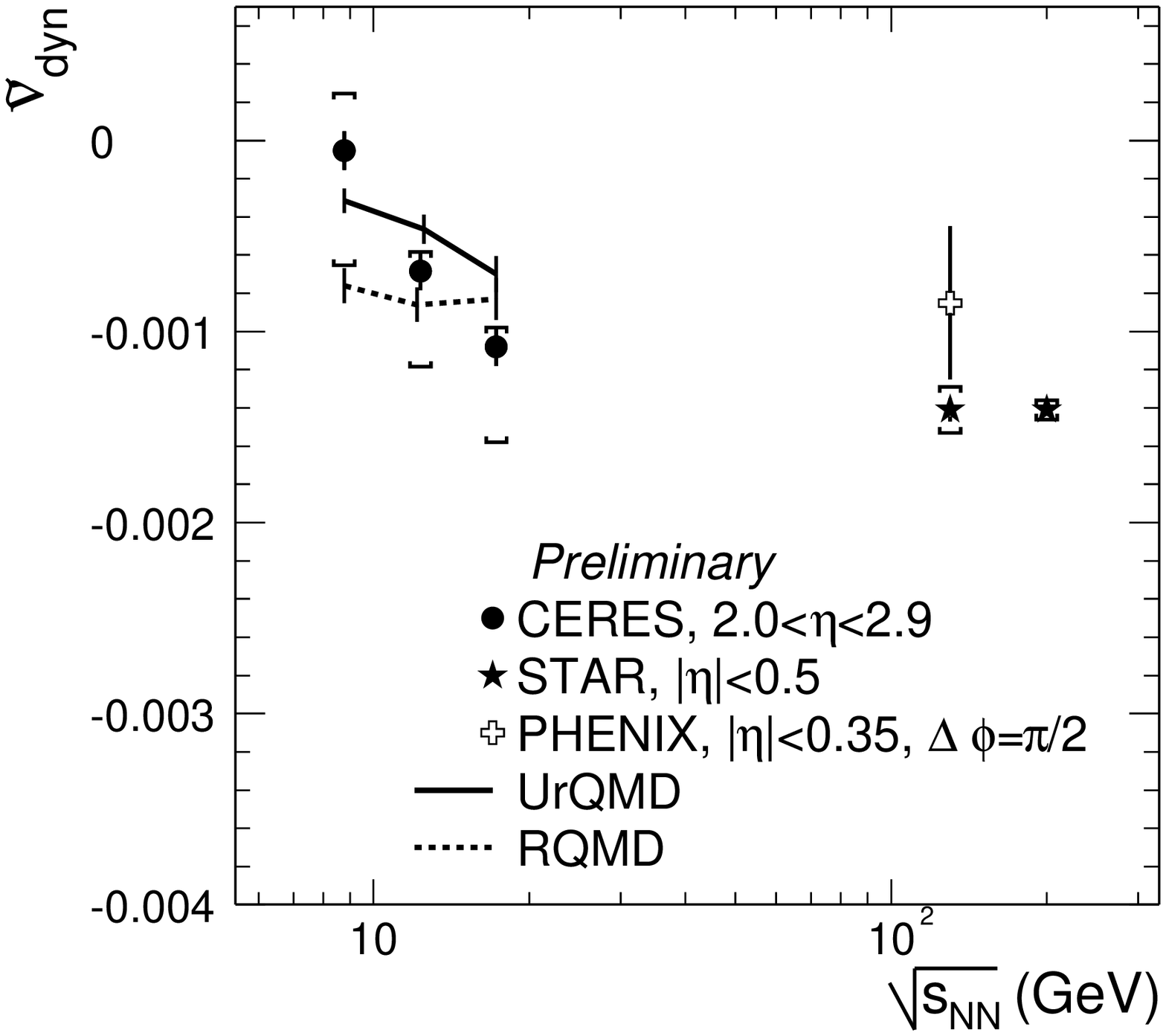,width=0.85\textwidth}}
\caption{$\tilde{\nu}_{dyn}$ as a
    function of {$\sqrt{s_{NN}}$} in central events.}
\label{Fig:vdyns}
\end{center}
\end{minipage}
\end{figure}

Fig.~\ref{Fig:nvdyn} shows $\nu_{dyn}$ multiplied with mean charged particle
multiplicity within the acceptance ($\langle N\rangle$) as a function of $N_{part}$ at the three
beam energies. 
The dashed lines show the charge conservation limit,
defined as $- 4 \langle N\rangle / \langle N\rangle_{4\pi}$, where
$\langle N\rangle_{4\pi}$ is the total charged particle multiplicity which is assumed
to scale with $\langle N\rangle$.
The $\langle N\rangle_{4\pi}$ is estimated based on the NA49
measurements~\cite{NA49:2002}. The observed fluctuations are lower than
the charge conservation limit.
Since the fluctuations are much above the QGP model predictions of $\sim-3.5$~\cite{Asakawa:2000,Jeon-Koch:2000}, no indication for the
formation of a deconfined phase is found.
Decreasing tendencies of $\langle N \rangle \nu_{dyn}$ with $N_{part}$ are
observed at the three beam energies. Whether this is due to a
deviation of the charge
conservation limit from the multiplicity scaling,
or due to a real deviation of the fluctuations
from the multiplicity scaling, is under investigation.

Fig.~\ref{Fig:vdyns} shows, as a function of $\sqrt{s_{NN}}$,
a compilation of CERES, STAR~\cite{STAR-q:2003,Pruneau:2003}, and
PHENIX results~\cite{PHENIX-q:2002} for $\nu_{dyn}$ corrected for
charge conservation ($\tilde{\nu}_{dyn}$). A decrease of $\tilde{\nu}_{dyn}$ is observed at SPS energies,
while a decrease is tiny from SPS top energy to RHIC energies.
Both RQMD and UrQMD are in reasonable agreement with the CERES data.


\section{Conclusions}
Dynamical mean $p_{T}$ fluctuations of $\sim 1$~\% in central
events are observed, which are similar to the fluctuations measured at
RHIC. The $F_{p_{T}}$ shows non-monotonic centrality dependence with a
maximum
of $\sim2.8$~\% in semi-central events, which is about a factor of 2 enhanced
over the p+p extrapolation. A similar tendency is observed also by
NA49, PHENIX, and STAR.
RQMD qualitatively reproduces both enhancement of
fluctuations in semi-central events and increase of mean $p_{T}$
as a function of centrality.

The observed dynamical net charge fluctuations are smaller than the charge
conservation limit. The $\langle N \rangle \nu_{dyn}$ slightly decreases from
peripheral to central events. The $\tilde{\nu}_{dyn}$ decreases at SPS
energies but varies little from the SPS top energy to RHIC energies.
UrQMD and RQMD reproduce the observed fluctuations at SPS.

\end{document}